\documentclass[fleqn,twoside]{article}
\usepackage{espcrc2}

\usepackage{graphicx}
\usepackage[figuresright]{rotating}
\usepackage{amssymb}

\newcommand{\beq}{\begin{equation}}
\newcommand{\eeq}{\end{equation}}
\newcommand{\ber}{\begin{eqnarray}}
\newcommand{\eer}{\end{eqnarray}}

\newcommand{\cD}{\cal{D}}

\newcommand{\AmS}{{\protect\the\textfont2
  A\kern-.1667em\lower.5ex\hbox{M}\kern-.125emS}}

\begin{document}

\title{
\vspace{-1.8cm}
\hfill \rm \null \hfill
 \hbox{\normalsize ADP-02-65/T505} \hskip1.5cm \phantom{?}\\
\vspace{+1.3cm}QCD, Gauge Fixing, and the Gribov Problem}

\author{A. G. Williams 
        \address{Special Research Centre for the Subatomic Structure of
                 Matter,
        Adelaide University, Adelaide 5005, Australia}}

\begin{abstract}
The standard techniques of gauge-fixing, such as covariant gauge
fixing, are entirely adequate for the purposes of studies of
perturbative QCD.  However, they fail in the nonperturbative regime
due to the presence of Gribov copies.
These copies arise because standard local gauge fixing methods
do not completely fix the gauge.  Known Gribov-copy-free gauges, such as
Laplacian gauge, are manifestly non-local.  These issues are examined and
the implications of non-local gauge-fixing for ghost fields, BRST
invariance, and the proof of renormalizability of QCD are considered.

\vspace{1pc}
\end{abstract}

\maketitle

\section{GAUGE FIXING}

The naive Lagrangian density of QCD is
\beq
{\cal L}_{\mathrm{QCD}} = -\frac{1}{4}F^{\mu\nu}F_{\mu\nu} 
                          + \sum_f{\bar{q}}_f (iD\!\!\!\!/-m_f)q_f,
\label{eq:LagDen}
\eeq
where the index $f$ corresponds to the quark flavours. The naive
Lagrangian is neither gauge-fixed nor renormalized, however it is
invariant under local $SU(3)_c$ gauge  transformations $g(x)$.
For arbitrary, small $\omega^a(x)$ we have
\beq
g(x) \equiv \exp\left\{-ig_s\left(\frac{\lambda^a}{2}\right)\omega^a(x)\right\}\in SU(3),
\eeq
where the $\lambda^a/2\equiv t^a$ are the generators of the gauge group
$SU(3)$ and the index $a$ runs over the eight generator labels
$a=1,2,...,8$.

Consider some gauge-invariant Green's function
(for the time being we shall
concern ourselves only with gluons)
\beq
\langle \Omega|\ T(\hat O[A])\ |\Omega\rangle 
 = \frac{\int{\cD}A\ O[A]\ e^{iS[A]}}{\int{\cD}A\ e^{iS[A]}} \, ,
\label{Eq:GreensF}
\eeq
where $O[A]$ is some gauge-independent quantity depending on the gauge
field, $A_\mu(x)$.
We see that the gauge-independence of $O[A]$ and $S[A]$ 
gives rise to an infinite quantity in both
the numerator and denominator,
which must be eliminated by gauge-fixing.  The Minkowski-space
Green's functions are defined as the Wick-rotated versions of the
Euclidean ones.

\begin{figure}
\includegraphics[angle=0,width=15pc]{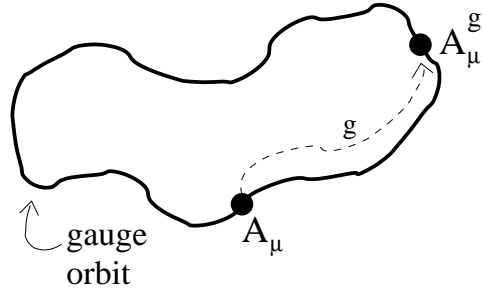}
\caption{Illustration of the gauge orbit containing $A_\mu$ and indicating
the effect of acting on $A_\mu$ with the gauge transformation $g$.  The
action $S[A]$ is constant around the orbit.}
\label{fig:Gorbit}
\end{figure}

The gauge orbit for some configuration $A_\mu$ is defined to be the set of
all gauge-equivalent configurations. Each point $A_\mu^g$ on the gauge
orbit is obtained by acting upon $A_\mu$ with the gauge transformation $g$.
By definition the action, $S[A]$, is gauge invariant and so all
configurations on the gauge orbit have the same action, e.g., see the
illustration in Fig.~\ref{fig:Gorbit}.

The integral over the gauge fields can be written as the integral
over a full set of gauge-inequivalent (i.e., gauge-fixed) configurations,
$\int{\cD}A^{\rm g.f.}$, and an integral over the gauge group 
$\int{\cD}g$.  In other words, $\int{\cD}A^{\rm g.f.}$ is an integral
{\em over the set of all possible gauge orbits} and $\int{\cD}g$ is an
integral {\em around the gauge orbits}.  Thus we can write
\beq
 \int{\cD}A \equiv  \int{\cD}A^{\rm g.f.}\int{\cD}g.
\eeq
To make integrals such as those in the numerator and denominator
of Eq.~(\ref{Eq:GreensF}) finite and also to study
gauge-dependent quantities in a meaningful way, we need to 
eliminate this integral around the gauge orbit, $\int{\cD}g$.

\section{GRIBOV COPIES AND THE FADDEEV-POPOV DETERMINANT}

\begin{figure}
\includegraphics[angle=0,width=15pc]{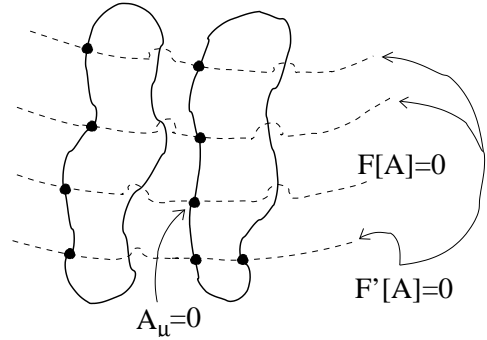}
\caption{Ideal, $F[A]$, and non-ideal, $F'[A]$, gauge-fixing.}
\label{fig:manyorbit}
\end{figure}

Any gauge-fixing procedure defines a surface in gauge-field
configuration space. Fig.~\ref{fig:manyorbit} is a depiction of these
surfaces represented as dashed lines intersecting the gauge orbits within 
this configuration space.   Of course, in general, the gauge orbits are
hypersurfaces and so are the gauge-fixing surfaces.  Any gauge-fixing
surface must, by definition, only intersect the gauge orbits at distinct
isolated points in configuration space.  For this reason, it is sufficient
to use lines for the simple illustration of the concepts here.
An ideal (or complete) gauge-fixing condition, $F[A]=0$, defines a surface
that intersects each gauge orbit once and only once and by convention
contains the trivial configuration $A_\mu=0$.  A non-ideal gauge-fixing
condition, $F'[A]=0$, defines a surface or surfaces which intersect the
gauge orbit more than once. These multiple intersections of the non-ideal
gauge fixing surface(s) with the gauge orbit are referred to as Gribov
copies\cite{Giusti:2001xf,vanBaal:1997gu,Neuberger:xz,Testa:1998az,Testa:1999qc,Alkofer:2000wg}
.  Lorentz gauge ($\partial_\mu A^\mu(x)=0$) for example, has many
Gribov copies per gauge orbit.  By definition an ideal gauge
fixing is free from Gribov copies.  We refer to the ideal gauge-fixing
surface $F[A]=0$ as the Fundamental Modular Region (FMR) for that gauge
choice.  Typically the gauge fixing condition depends on a space-time
coordinate, (e.g., Lorentz gauge, axial gauge, {\em etc.}), and so we
write the gauge fixing condition more generally as $F([A];x)=0$. 

Let us denote one arbitrary gauge configuration per gauge orbit as
$A_\mu^0$ and let this correspond to the ``origin'' of gauge
configurations on that gauge orbit, i.e., to $g=0$ on that orbit.
Then each gauge orbit can be labelled by $A_\mu^0$ and the set of all
such $A_\mu^0$ is equivalent to one particular, complete specification of
the gauge.
Under a gauge transformation, $g$, we move from the origin of the gauge
orbit to the configuration, $A_\mu^g$, where by definition
$A_\mu^0 \stackrel{g}{\longrightarrow} A_\mu^g = 
    g A_\mu^0 g^\dagger - (i/g_s)(\partial_\mu g)g^\dagger$.
Let us denote for each gauge orbit the gauge transformation,
$\tilde g\equiv \tilde g[A^0]$, as the transformation which
takes us from the origin of that
orbit, $A_\mu^0$, to the configuration, $A_\mu^{\rm g.f.}$,
which lies on the ideal gauge-fixed surface specified by $F([A];x)=0$.
In other words, we have 
$ F([A];x)|_{A^{\tilde g}}=0$ for 
$A^{\tilde g}\equiv A_\mu^{\rm g.f.}\in {\mathrm {FMR}}$.

The {\em inverse Faddeev-Popov determinant} is defined as the integral over
the gauge group of the gauge-fixing condition, i.e.,
\ber
\Delta^{-1}_F[A^{\rm g.f.}] & \equiv& \int{\cD}g\ \delta[F[A]] \nonumber \\ 
                        && \hskip-2cm =  \int{\cD}g\ \delta(g-{\tilde{g}}) 
   \left|\det\left(\frac{\delta F([A];x)}{\delta g(y)}\right)
                              \right|^{-1} \, . 
\eer
Let us define the matrix $M_F[A]$ as
\beq
M_F([A];x,y)^{ab}\equiv\frac{\delta F^a([A];x)}{\delta g^b(y)} \, .
\eeq
Then the {\em Faddeev-Popov determinant} for an arbitrary configuration
$A_\mu$ can be defined as
$\Delta_F[A] \equiv \left|\det M_F[A]\right|$.
(The reason for the name is now clear).  Note that we
have consistency, since
$\Delta^{-1}_F[A^{\rm g.f.}] \equiv \Delta^{-1}_F[A^{\tilde g}] =
\int{\cD}g\ \delta(g-{\tilde{g}})\Delta^{-1}_F[A]$.

We have $1 = \int{\cD}g\ \Delta_F[A]\,\, \delta[F[A]]$
by definition and hence
\ber
  \int{\cD}A^{\rm g.f.} \hskip-0.25cm &\equiv&  
          \int{\cD}A^{\rm g.f.}\!\!\int{\cD}g\ \Delta_F[A]\,\, \delta[F[A]]
          \nonumber \\
             & = & \int{\cD}A\ \Delta_F[A]\,\, \delta[F[A]]
\eer
Since for an ideal gauge-fixing there is one and only one $\tilde{g}$
per gauge orbit, such that
$F([A];x)|_{\tilde g}=0$, then $|$det$M_F[A]|$ is non-zero on the FMR. 
It follows that if there is at least one smooth path between any two
configurations in the FMR and since the determinant cannot be zero on the 
FMR, then it cannot change sign on the FMR. The {\em Gribov horizon} is
defined to by those configurations with $\det M_F[A]=0$ which lie
closest to the FMR.  By
definition the determinant can change sign on or outside this horizon.
Clearly,
the FMR is contained within the Gribov horizon and for an ideal gauge
fixing, since the sign of the determinant cannot change, we can replace
$|\det M_F|$ with $\det M_F$, [i.e., the overall sign of the functional
integral is normalized away in Eq.~(\ref{Eq:GreensF})].

These results are generalizations of results from ordinary calculus,
where
\beq
\left|{\mathrm {det}}\left(\frac{\partial f_i}{\partial x_j}\right)
        \right|_{\vec{f}=0}^{-1}
           = \int dx_1\cdots dx_n\, \delta^{(n)}({\vec{f}}({\vec{x}}))
\eeq
and if there is one and only one ${\vec{x}}$ which is a solution of
${\vec{f}}({\vec{x}})=0$ then the matrix
$M_{ij} \equiv \partial f_i/\partial x_j$
is invertible (i.e., non-singular) on the hypersurface 
${\vec{f}}({\vec{x}})=0$ and hence $\det M\neq 0$. 

\section{GENERALIZED FADDEEV-POPOV TECHNIQUE}
Let us now assume that we have a family of {\em ideal}
gauge fixings $F([A];x)=f([A];x)-c(x)$
for any Lorentz scalar $c(x)$ and for $f([A];x)$ being some Lorentz scalar
function,
(e.g., $\partial^\mu A_\mu(x)$ or $n^\mu A_\mu(x)$ or similar or any
nonlocal generalizations of these).
Therefore, using the fact that we remain in the FMR and can drop the
modulus on the determinant, we have
\beq
\int{\cD}A^{\rm g.f.} = \int{\cD}A\ \det M_F[A]\ \delta[f[A]-c]
     \, .
\eeq
Since $c(x)$ is an arbitrary function, we can define a new
``gauge'' as the Gaussian weighted average over $c(x)$, i.e.,
\begin{eqnarray}
\int{\cD}A^{\rm g.f.} \hskip-0.25cm
                  &\propto & \int{\cD}c\ {\mathrm{exp}}\left\{
                      -\frac{i}{2\xi}\int d^4x c(x)^2 \right\} \nonumber \\
                  &  &\times \int{\cD}A\ {\mathrm {det}} M_F[A]\
                                   \delta[f[A]-c] \nonumber \\
                  & \propto & \int{\cD}A\ {\mathrm {det}}M_F[A] \nonumber \\
                  &  & \times {\mathrm{exp}}\left\{
                      -\frac{i}{2\xi}\int d^4x f([A];x)^2 \right\} \nonumber \\
                  & \propto & \int{\cD}A {\cD}\chi{\cD}{\bar{\chi}}\ \ 
                      {\mathrm{exp}}\left\{
                      -i\int d^4xd^4y\right. \nonumber \\
                  & & \left. \frac{}{}\times 
                     {\bar{\chi}}(x)M_F([A];x,y)\chi(y)\right\} \nonumber \\
                  & & \times {\mathrm{exp}}\left\{
                      -\frac{i}{2\xi}\int d^4x f([A];x)^2 \right\} \, ,
\end{eqnarray}
where we have introduced the anti-commuting ghost fields
$\chi$ and $\bar{\chi}$.
Note that this kind of ideal gauge fixing does not choose just one gauge
configuration on the gauge orbit, but rather is some Gaussian
weighted average over gauge fields on the gauge orbit.
We then obtain
\ber
\langle\Omega|\ T(\hat{O}[...])\ |\Omega\rangle & = & \nonumber \\
 & & \hskip-2cm
    \frac{\int{\cD}q{\cD}\bar{q}{\cD}A{\cD}\chi{\cD}{\bar{\chi}}\
        O[...]\ e^{iS_{\xi}[...]}}
         {\int{\cD}q{\cD}\bar{q}{\cD}A{\cD}\chi{\cD}{\bar{\chi}}\ 
        e^{iS_{\xi}[...]}}\, , 
\eer
where 
\begin{eqnarray}
S_{\xi}[q,\bar{q},A,\chi,{\bar{\chi}}] & = & \int d^4x \left[
   -\frac{1}{4}F^{a\mu\nu}F^a_{\mu\nu} \right. \nonumber \\
   & & \hskip-2.5cm -\frac{1}{2\xi}\left(f([A];x)\right)^2  
         \left. + \sum_f{\bar{q}}_f (iD\!\!\!\!/-m_f)q_f\right]  \nonumber \\
   & & \hskip-2.5cm + \int d^4xd^4y\ 
             {\bar{\chi}}(x)M_F([A];x,y)\chi(y) \, .
\label{Eq:gen_action}
\end{eqnarray} 

\section{STANDARD GAUGE FIXING}

  We can now recover standard gauge fixing schemes as special cases
of this generalized form.  First consider standard covariant
gauge, which we obtain by taking
$f([A];x)=\partial_\mu A^{\mu}(x)$ and by neglecting the fact that this
leads to Gribov copies.
We need to evaluate $M_F[A]$ in the vicinity of the gauge-fixing surface
for this choice:
\begin{eqnarray}
M_F([A];x,y)^{ab} &=& \frac{\delta F^a([A];x)}{\delta g^b(y)}
                               \nonumber \\
      && \hskip-2.5cm
        = \frac{\delta [\partial_\mu A^{a\mu}(x)-c(x)]}{\delta g^b(y)} 
      = \partial^x_\mu\frac{\delta A^{a\mu}(x)}{\delta g^b(y)]} \, .
\end{eqnarray}
Under an infinitesimal gauge transformation $g$
we have
\ber
A_\mu^a(x) &\stackrel{{\rm small}\ g}{\longrightarrow}&
  (A^g)_\mu^a(x) = A_\mu^a(x) \nonumber \\
 && \hskip-2cm + g_sf^{abc}\omega^b(x)A_\mu^c(x)
                -\partial_\mu\omega^a(x) + {\cal O}(\omega^2)
\eer
and hence in the neighbourhood of the gauge fixing surface (i.e., for
small fluctations along the gauge orbit around $A_\mu^{\rm g.f.}$), we have
\begin{eqnarray}
M_F([A];x,y)^{ab} \hskip-0.1cm
   &=& \hskip-0.1cm \left. \partial^x_\mu
     \frac{\delta A^{a\mu}(x)}{\delta \omega^b(y)]} \right|_{\omega=0}
       \\
   && \hskip-3.2cm 
        = \partial^x_\mu \left(\frac{}{}[-\partial^{x\mu}\delta^{ab} 
               + g_s f^{abc}A^{c\mu}(x)] 
                \times \delta^{(4)}(x-y)\right) \, . \nonumber 
\end{eqnarray}                                 
We then recover the standard covariant gauge-fixed form of the QCD action
\begin{eqnarray}
S_{\xi}[q,\bar{q},A,\chi,{\bar{\chi}}]
  & = & \int d^4x \left[
   -\frac{1}{4}F^{a\mu\nu}F^a_{\mu\nu} \right. \nonumber \\
   && \hskip-2.5cm 
       \left. -\frac{1}{2\xi}\left(\partial_\mu A^{\mu}(x)\right)^2 
         + \sum_f{\bar{q}}_f (iD\!\!\!\!/-m_f)q_f\right] \nonumber \\
   && \hskip-2.5cm 
      + (\partial_\mu\bar{\chi}_a)(\partial^\mu\delta^{ab}
       -gf_{abc}A_c^\mu)\chi_b \, .
\end{eqnarray}
However, this gauge fixing has not removed the Gribov copies and so
the formal manipulations which lead to this action are not valid. 
This Lorentz covariant set of naive gauges corresponds to a Gaussian
weighted average over generalized Lorentz gauges, where the gauge parameter
$\xi$ is the width of the Gaussian distribution over the configurations
on the gauge orbit.
Setting $\xi=0$ we see that the width vanishes and we obtain Landau 
gauge (equivalent to Lorentz gauge, $\partial^\mu A_\mu(x)=0$).
Choosing $\xi=1$ is referred to as ``Feynman gauge'' and so on.

We can similarly recover the standard QCD action for axial gauge, where
$n_\mu A^\mu(x)=0$.
Proceeding as for the generalized covariant gauge, we first identify
$f([A];x)=n_\mu A^\mu(x)$ and obtain the gauge-fixed action
\begin{eqnarray}
S_{\xi}[q,\bar{q},A]&=&\int d^4x 
     \left[-\frac{1}{4}F^{a\mu\nu}F^a_{\mu\nu} \right. \nonumber \\
  & & \left. \hskip-2cm
       -\frac{1}{2\xi}\left(n_\mu A^\mu(x)\right)^2
       + \sum_f{\bar{q}}_f (iD\!\!\!\!/-m_f)q_f\right] \, .
\end{eqnarray}
Taking the ``Landau-like'' zero-width limit $\xi \rightarrow 0$ we
select $n_\mu A^\mu(x)=0$ exactly and recover the usual axial-gauge
fixed QCD action.  Axial gauge does not involve ghost fields, since
in this case
\ber
M_F([A^{\rm g.f.}];x,y)^{ab}
    &=& \left. n_\mu\frac{\delta A^{a\mu}(x)}
         {\delta \omega^b(y)]} \right|_{\omega=0} \nonumber \\
    && \hskip-2cm
      = n_\mu \left([-\partial^{x\mu}\delta^{ab}]\delta^{(4)}(x-y)\right)
         \, ,
\eer
which is independent of $A_\mu$ since $n^\mu A_\mu^{\rm g.f.}(x)=0$. 
In other words, the gauge field does not appear in $M_F[A]$ on the
gauge-fixed surface.
Unfortunately axial gauge suffers from singularities which lead to
significant difficulties when trying to define perturbation theory
beyond one loop. A related feature is that axial gauge is
not a complete gauge fixing prescription.  While there are complete
versions of axial gauge on the lattice, these always involve a nonlocal
element, or reintroduce Gribov copies at the boundary so as not to
destroy the Polyakov loop.

\section{DISCUSSION AND CONCLUSIONS}

There is no known Gribov-copy-free gauge fixing which is a {\em local}
function of $A_{\mu}(x)$.  In other words, such a gauge fixing cannot
be expressed as a function of $A_\mu(x)$ and a finite number of its
derivatives, i.e., $F([A];x)\neq F(\partial_\mu,A_\mu(x))$ for all $x$.
Hence, the gauge-fixed action, $S_{\xi}[\cdots]$, in Eq.~(\ref{Eq:gen_action})
becomes non-local and gives rise to a nonlocal quantum field theory.
Since this action serves as the basis for the proof of the renormalizability
of QCD, the proof of asymptotic freedom, local BRS symmetry, and the
Schwinger-Dyson equations (to name but a few) the nonlocality of the action
leaves us without a reliable basis from which to prove these features of
QCD in the nonperturbative context.

It is well-established that QCD is asymptotically free, i.e., it is
weak-coupling at large momenta.  In the weak coupling limit the functional
integral is dominated by small action configurations. As a consequence,
momentum-space Green's functions at large momenta will receive their
dominant contributions in the path integral from configurations near the
trivial gauge orbit, i.e., the orbit containing $A_\mu=0$, since this
orbit minimizes the action.  If we use standard
gauge fixing, which neglects the fact that Gribov copies are present, then
at large momenta $\int{\cD}A$ will be dominated by configurations lying
on the gauge-fixed surfaces in the neighbourhood of {\em each} of the Gribov
copies on the trivial orbit.  Since for small field fluctuations the Gribov
copies cannot be aware of each other, we merely overcount the contribution
by a factor equal to the number of copies on the trivial orbit.  This
overcounting is normalized away by the ratio in Eq.~(\ref{Eq:GreensF})
and becomes irrelevant.  Thus it is possible to understand why Gribov
copies can be neglected at large momenta and why it is sufficient to
use standard gauge fixing schemes as the basis for calculations in
perturbative QCD.  Since renormalizability is an ultraviolet issue,
there is no question about the renormalizability of QCD.

Lattice QCD has provided numerical confirmation of asymptotic freedom, 
so let us now turn our attention to the matter of Gribov copies in lattice
QCD.  Since the observable $O[A]$ and the action are both gauge-invariant
it doesn't matter whether we sample from the FMR of an ideal gauge-fixing
or elsewhere on the gauge orbit.  The trick is simply to sample 
{\em at most once} from each orbit.  Since there are an infinite number of
gauge orbits (even  on the lattice), no finite ensemble will ever sample
the same orbit twice.  This makes Gribov copies and gauge-fixing irrelevant in
the calculation of color-singlet quantities on the lattice.

The calculation of gauge-{\em dependent} Green's functions on the lattice does
require that the gauge be fixed.  The standard choice is naive lattice
Landau gauge, which selects essentially at random between the Landau
gauge Gribov copies for the gauge orbits represented in the ensemble.
This means that, while the gauge fixing is well-defined in that there
are no Gribov copies, the Landau gauge-fixed configurations are not from
a single connected FMR.  For this reason lattice studies of gluon and quark
propagators are now being extended to Laplacian gauge for comparison.
Laplacian gauge is interesting because it is Gribov-copy-free (except
on a set of configurations of measure zero) and it reduces to Landau
gauge at large momenta.  Lattice calculations of the Laplacian gauge and
Landau gauge quark and gluon propagators converge at large momenta and hence
are consistent with this expectation.

In conclusion, it should be noted that throughout this discussion there
has been the implicit assumption that nonperturbative QCD should be
defined in such a way that each gauge orbit is represented only once
in the functional integral, i.e., that it be defined to have no Gribov
copies.  This is the definition of nonpertubative QCD implicitly assumed
in lattice QCD studies.
We have seen that this assumption destroys locality and the
BRS invariance of the theory.  An equally valid point of view is that
locality and BRS symmetry are central to the definition of QCD and must
not be sacrificed in the nonpertubative regime, (see, e.g., 
Ref.~\cite{Neuberger:xz,Testa:1998az,Testa:1999qc,Alkofer:2000wg}).
This viewpoint
impies that Gribov copies are necessarily present, that gauge orbits
are multiply represented, and that the definition of nonperturbative QCD
must be considered with some care.  Since these nonpertubative definitions
of QCD appear to be different, establishing which is the one appropriate
for the description of the physical world is of considerable importance.

\end{document}